\documentclass{article}
\usepackage{arxiv}

\usepackage{mathtools}
\usepackage{amsfonts}
\usepackage{graphicx, subcaption}
\usepackage{algorithm2e}
\usepackage{url}
\usepackage{url}

\begin{document}
\title{Inferring probabilistic Boolean networks from steady-state gene data samples}


\author{Vytenis Šliogeris, Leandros Maglaras, Sotiris Moschoyiannis
\thanks{Vytenis Šliogeris and Sotiris Moschoyiannis are with the School of Computer Science \& Electronic Engineering, University of Surrey, GU2 7XH, UK (e-mail: s.moschoyiannis@surrey.ac.uk, v.sliogeris@surrey.ac.uk). Sotiris Moschoyiannis and Vytenis Sliogeris have been partly funded by UKRI Innovate UK, grant 77032. Leandros Maglaras is with the School of Computer Science and Informatics, De Montfort University,
Leicester, UK (e-mail: leandros.maglaras@dmu.ac.uk).}
}

\maketitle

\begin{abstract}
	Probabilistic Boolean Networks have been proposed for estimating the behaviour of dynamical systems as they combine rule-based modelling with uncertainty principles. Inferring PBNs directly from gene data is challenging however, especially when data is costly to collect and/or noisy, e.g., in the case of gene expression profile data. In this paper, we present a reproducible method for inferring PBNs directly from {\it real} gene expression data measurements taken when the system was at a steady state. The steady-state dynamics of PBNs is of special interest in the analysis of biological machinery. The proposed approach does not rely on reconstructing the state evolution of the network, which is computationally intractable for larger networks. We demonstrate the method on samples of real gene expression profiling data from a well-known study on metastatic melanoma. The pipeline is implemented using Python and we make it publicly available. 
	
\end{abstract}

\section{Introduction}

Rapid progress in the development of next-generation sequencing technologies for genomics has provided valuable insights into complex biological systems \cite{single-seq-advances-2016}. Modelling single-cell or gene networks is becoming increasingly important. The question of modelling complex molecular regulatory networks is an important one for bioinformatics. The goal of systems biology is to intervene on the state of the cell, using the dynamics of the underlying regulatory network. A model that could accurately represent such dynamics could be used for analysis, including control \cite{liu-2011-controllability,karlsen-2018-evolution,papagiannis19r,wu-2019-policy-iteration-stochastic,CN2020-drl-pbn}, steady-state distribution \cite{ssd-shmulevich-2003,mos-shi-FI2004,ching-2007-ssd-analysis,kobayashi-2017-infer-pbn}, observability \cite{JLu-observability-2019,observability-kuize-2020,CN2017-impact-of-removing}. Such analyses aid the development of genetic 
therapies \cite{fumia-2013-gene-therapies}. 

Boolean Networks (BNs) were introduced for this purpose by  Kauffman \cite{kauffman69}. 
In brief, a BN comprises a set of Boolean variables, each variable representing the on/off state of a gene, while interactions between genes are expressed by Boolean functions. It was found that even randomly generated BNs exhibit behaviour reminiscent of gene regulatory networks, with naturally arising attractor states which represent cell types or the phenotype \cite{voukantsis2019microcell,MAPK-nw-analysis-2022}. This explains the popularity of BNs for modelling gene interactions \cite{albert-2003-topology,davidich-BN-models-in-biology}.

However, with few exceptions, gene expression data suggests a number of possible successor states to any given state in a BN, thereby refuting the determinism inherent in BNs.  
Thus, a {\it probabilistic} BN (PBN) was introduced by Shmulevich et al. \cite{smulevich02} in which the definition of a BN was adapted such that for each gene, at each time point, a Boolean function (and predictor gene set) is chosen with some conditional probability \cite{shmulevich10}. 

Inferring the PBN representation of a gene regulatory network (GRN) is quite involved. First, the directed graph expressing interactions between genes needs to be constructed; then, the Boolean functions need to be determined; followed by determining the probabilities of selecting a Boolean function as well as the number of candidate functions on each gene. Existing work (cf Section \ref{relWork}) tends to focus on inference from {\it time-series} gene expression data as the temporal aspect reveals the transition structure of the corresponding PBN. However, as already pointed out in \cite{joseph-2004-time-series-gene-data}, there are concerns over the number of (typically expensive to obtain) observations needed in such gene microarray data. Approaches based on ODEs (e.g., \cite{matsumoto2017scode}) require lots of observations to tune the large number of parameters of the model, while in practice only a handful are available. More such observations are available when the underlying gene network is at a {\it steady state} \cite{ssd-shmulevich-2003}, e.g, see gene expression profiles of melanoma by Bittner {\it et al} \cite{bittner00}. 


In this paper, we propose a systematic method for inferring PBNs directly from real gene expression data measurements, collected using microarray technology, when the system is at a {\it steady-state}. The steady-state (long-run) behaviour of a PBN is of interest to system biology as it allows to determine the long-term influence of a gene on another gene or determine the long-term joint probabilistic behaviour of a few selected genes \cite{ssd-shmulevich-2003}. 

The key contribution of our paper is a reproducible pipeline for going from gene (steady-state) data samples to the PBN representation of the long-run behaviour of the underlying genetic network. We use a predictor gene set rather than temporal data to infer the "transition structure". Unlike other proposals, our method does not require the construction of the {\it probability transition matrix}, whose size grows exponentially on the number of nodes, and hence becomes computationally intractable for larger networks \cite{akutsu-2007-control}.

The remainder of the paper is structured as follows. Section \ref{relWork} outlines related work. Preliminary background knowledge is presented in Section \ref{BG}. The main algorithm for our inference method is in Section \ref{process}. PBNs are produced in Section \ref{eval} using the process described in Section \ref{analysis}.
Concluding remarks are in Section \ref{conclusion}.

\section{Related Work}
\label{relWork}

There have been various methods for PBN inference, focusing on causality, using different types of gene data \cite{survey-on-causal-models}. Previous work on PBN inference from time series gene data includes \cite{temp-BN-inference-2001}, SCODE \cite{matsumoto2017scode} with ODEs, and most recently the Stochastic Conjunctive Normal Form (SCNF) -based method by Apostolopoulou {\it et al} \cite{apostolo-2019-tractable} which can address larger networks.  

Previous work on inference from steady-state data samples is relatively limited and goes back to Shmulevich {\it et al} \cite{ssd-shmulevich-2003}. A tool for computing the ssd probabilities has been proposed in \cite{assa-pbn-analyser-2015}. Melkman {\it et al} \cite{melkman-PBTNs-2018} infer {\it threshold} PBNs, a particular version of PBNs where every input threshold function of a node must have the same number of parameters and also satisfy certain stringent conditions. Kobayashi {\it et al} \cite{kobayashi-2017-infer-pbn} construct PBNs from BNs by casting inference as an integer linear programming problem and constructs a PBN that fits the given steady-state distribution. 

Kim et al. \cite{kim03} use steady-state gene data samples from the study on {\it metastatic melanoma} by Bittner et al. \cite{bittner00} (we use the same data here). They choose the genes for their PBN using a combination of Coefficient of Determination (COD) analysis and biological background knowledge (we do not assume any prior knowledge). 
For the functions, they ternarise their data, and construct Lookup Tables in place of the functions for each gene. They also analyse the PBNs produced by analysing the steady-state distribution (ssd) of the resulting network. 

Shmulevich et al. \cite{smulevich02}, who introduced PBNs, describe a method for determining functions for nodes in a PBN. This requires finding sets of input genes which have high COD with the target gene, and using the predictive model used for the calculation of the COD as the function for the particular set of input genes. The probability for choosing the particular input gene set is proportional to the COD of the input gene set.

Since discretisation of gene data is an important factor for inference. Chen et al. \cite{chen97} describe a method for quantising gene data using the expressions of {\it housekeeping} genes within the dataset. Housekeeping genes are genes which keep a constant expression, as they perform important functions within the cell. Since they have a constant expression, they can be used to estimate the probability distribution function (PDF) of the gene expressions within a microarray.
The constructed PDF can be used for using a hypothesis test to determine whether or not a gene is over- or under-expressed.
However, this method hinges on knowledge of which of the genes are housekeeping genes and this typically is not readily available. 


As discussed in the introductory section, we focus on constructing PBNs from real, microarray gene data samples, collected while the system is in a sterady-state, instead of simulated, time-series data or starting from BNs. We present a reproducible method to perform such a task.

\section{Preliminaries}
\label{BG}

\subsection{Boolean Networks}

A BN \cite{kauffman69} is a directed graph, $G = \{V,E\}$, comprised of vertices $V$ and edges $E$.
The vertices $v \in V$ represent the Boolean variables, which in this case represent genes in a gene regulatory network. 
The directed edges $\{v_i, v_j\} = e_{i,j} \in E$ represent that one variable, $v_i$, influences another, $v_j$.
Each vertex is associated with a Boolean function $f_i$ given by $f_i: \{0,1\}^{n_{in}} \mapsto \{0,1\}$.
The input for $f_i$ is a Boolean vector of length  $n_{in}$, which represents the states of all of the input vertices, and the output is a single Boolean value, which is then used as the next state of the variable $v_i$.
For a vertex $i$, the {\it input} vertices are the vertices from which all incoming edges originate, given by $\{v_j | \exists \{v_j, v_i\}\} = e_{j,i} \in E$. 

\subsection{Probabilistic Boolean Networks}

Probabilistic Boolean networks are an extension of Boolean networks.
They are directed graphs $G$, as in Boolean networks, except each function $f_i$ for each node $i$ in the case of Boolean networks is replaced by a set of Boolean functions $F_i = \{f_i^1, f_i^2, \dots, f_i^{l_i}\}$, and probabilities $c_i = \{c_i^1, c_i^2, \dots, c_i^{l_i}\}$. Hence, the logical function $f_i$ has $l_i$ possibilities, each with a corresponding conditional probability of being selected at every time step. 

More formally, during run time, a function $f_i^j$ for the node $v_i$ is chosen with probability $c_i^j$, $j \in [1,l_i]$.
PBNs are an extension to BNs in the sense that if each node within a PBN has a single function, it becomes identical to the BN.

\subsection{State Transition Graphs} 
\label{stg-ssd}

For each PBN there exists a state transition graph (STG).
An STG is a directed graph $G = \{V,E\}$, where the vertices $v_i \in V$ represent the possible states of the PBN, and the edges $\{v_i, v_j\} = e_{i,j} \in E$ represent the possibility of a transition from state $v_i$ to $v_j$.
Since the probability of getting to another state $v_j$ only depends on the current state $v_i$, we can say that the STG is a Markov chain.

By saying that the PBN has a steady state distribution (SSD), we mean that the STG of the PBN has a steady state distribution.
For an STG to have an SSD, it needs to be {\it ergodic} - that is, every state can reached from every other state. To guarantee that the STG is ergodic, random perturbations with low probability are introduced to the PBN.

\subsection{Microarray Gene Data Samples}

The data used to infer a PBN in our work was taken from the study of metastatic melanoma found in Bittner {\it et al} \cite{bittner00}, which has been extensively studied in the literature \cite{kim03,pal06,sirin13,CN2020-drl-pbn}.
The study extracts and analyses the gene expression profiles of 31 melanoma cells using microarray technology. To make sure that the gene expression levels used in inferring the corresponding PBN are those of genes when the network is in a steady state, the Kolmogorov-Smirnov (KS) statistic is applied, as discussed in more detail in Section \ref{analysis}. 

To utilise a particular gene in DNA, see \cite{chen97}, assuming the cell is at a steady-state, the relevant segment of the molecule must first be transcribed, producing messenger RNA (mRNA) which is accessible to the rest of the proteins.
The quantity of mRNA in a cell signifies the degree of protein production associated with a particular gene.

DNA microarrays measure the presence of mRNA within a cell.
The microarrays consist of a surface with an array of robotically placed complementary DNA for the genes to be analysed.
mRNA tightly bonds with complementary DNA, hence the microarray can be used to isolate different mRNA molecules.
The process is known as {\it hybridisation}.

The quantity of mRNA within a cell is measured by tagging the mRNA with fluorescent molecules, hybridising them with a microarray, and exciting the fluorescent molecules.
The emitted brightness is proportional to the amount of mRNA present.

Since the amount of mRNA differs depending on the gene, the data is normalised by dividing the values recorded by the values recorded from a reference probe. Since values recorded are non-negative, the ratio values are in the range of $[0,\infty)$. Furthermore, since we would expect the values of within the reference probe and the sample to not be different, the median for the ratio values is expected to be 1.
These are the values provided by Bittner et al. \cite{bittner00} in the form of a matrix of size 8,150 (number of genes) by 31 (number of samples). A small sample of the raw data is shown later in Fig. \ref{input-output}(a). 

For demonstrating our method of inferring a PBN, we work with the subset of melanoma genes analysed by Datta et al. \cite{datta03}, which are extensively studied in the literature \cite{kim03,pal06,kobayashi-2017-infer-pbn,sirin13,CN2020-drl-pbn}, namely {\it WNT5A, pirin, S100P, RET1, MART1, HADHB} and {\it STC2}. This offers straightforward validation for our approach since it produces the same PBN. 

It is worth noting that larger PBNs may be constructed following the pipeline described in this paper, and we have constructed the 28 node PBN given in \cite{sirin13} as well as 70 node PBN which includes the 28 nodes already studied in \cite{sirin13} padded with the 42 nodes with the highest weighting of importance, using discriminative weights, which determine how a gene changes during the experiment compared to the control cells \cite{bittner00}].

\subsection{Coefficient of Determination}

Coefficients of Determination (CODs) were described by Kim et al. \cite{kim00} as a method to determine which gene determines the state of which other gene.
A COD of a target variable, $Y$, with regards to an input variable, $X$, is a measure on how well the target variable can be predicted using the input variable.
A predictive model $f$ is used to predict the value of the target variable with and without the input variable, and compute the errors $\bar{e}$ and $e$ respectively.
The relative change of error of the predictive model is the COD $\theta$, given by eq. \ref{CODEq}:

\begin{equation}
\label{CODEq}
	\theta = \frac{\bar{e} - e}{\bar{e}}
\end{equation}

There are no constraints on what can be used as a predictive model. We opted for a perceptron.
This is because there exists a closed-form solution for linear regression of the perceptron, described by Kim et al. \cite{kim00}, which can be used instead of training. This aids in lowering the computation time.

The weights of a perceptron, $A$, can be computed using the closed form solution:
\begin{equation}
\label{CFEq}
	\begin{split}
		A &= R^+ \cdot C \\
		R &= X \cdot X^T \\
		C &= X \cdot Y \\
	\end{split}
\end{equation}
\subsection{Discretisation}
\label{discretisation}

Since PBNs use discrete values, the gene data which consists of real values has to be discretised.
Discretisation is a process where values get mapped from the real value domain to the integer domain.
For the problem at hand, since genes can be in one of two states, the range of the function should be either 0 or 1. Hence the function should take the form of:
\begin{equation}
	f: G \to G_d, x \geq 0, \forall x \in G, y \in \{0, 1\} \forall y \in G_d
\end{equation}
Such a method is described in detail in \cite{velarde08}.
It consists of deciding upon a threshold value $t$ with which all real values are compared. Each value then gets mapped to 0 if it is below the threshold, and to 1 otherwise, as given by eq. \ref{discEq}.
\begin{equation}
\label{discEq}
	G_d[x,y] = \left\{
	\begin{array}{ll}
		0 & G[x,y] < t\\
		1 & G[x,y] \geq t\\
	\end{array}
	\right.
\end{equation}
The threshold may be any metric.
Common metrics are means or medians.
The threshold may also be the boundary between the top $x\%$ of entries and the rest.

Shmulevich et al. \cite{shmulevich10} describe a process of using k-means clustering to cluster the data, and assigning values to the data points depending on the cluster they belong to.
However, since half of the data points lie in the range $(0,1)$, and the other half is in the range $(1, \infty)$, the lower cluster ends up larger, resulting in a larger threshold that produces more zeros.
This can be remedied by performing k-means clustering on the logarithms of the data points.
This makes the ranges of both halves the same, producing more representative clusters.

\section{Inference of PBNs}
\label{process}

\begin{figure}
	\centering
	\caption{Input and output for the inference method.}
	\label{input-output}
	\begin{subfigure}[t]{.55\linewidth}
        \includegraphics[width=\linewidth]{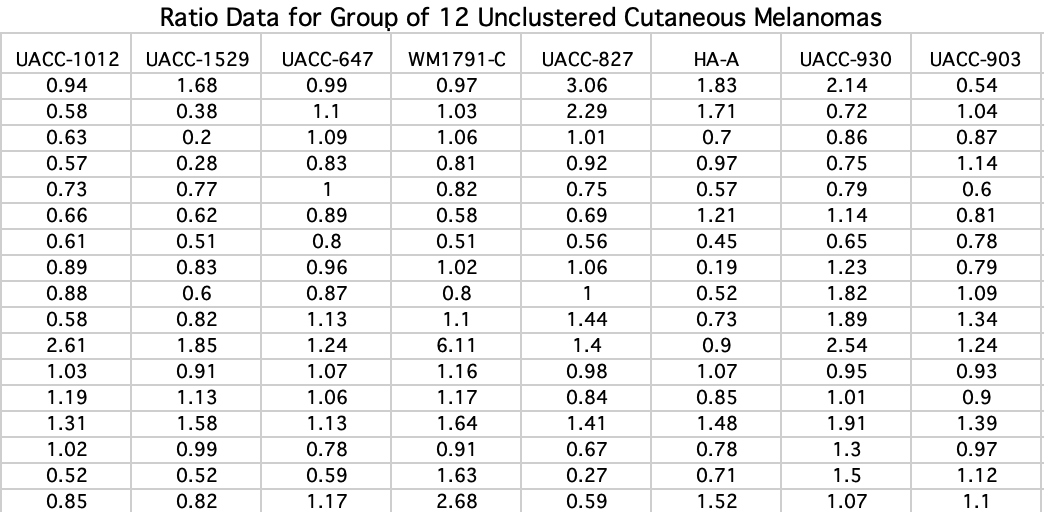}
        \caption{The input, gene expression values for 31 samples taken at a steady-state, given in the form of a matrix}
	\end{subfigure}
	\begin{subfigure}[t]{.4\linewidth}
        \includegraphics[width=\linewidth]{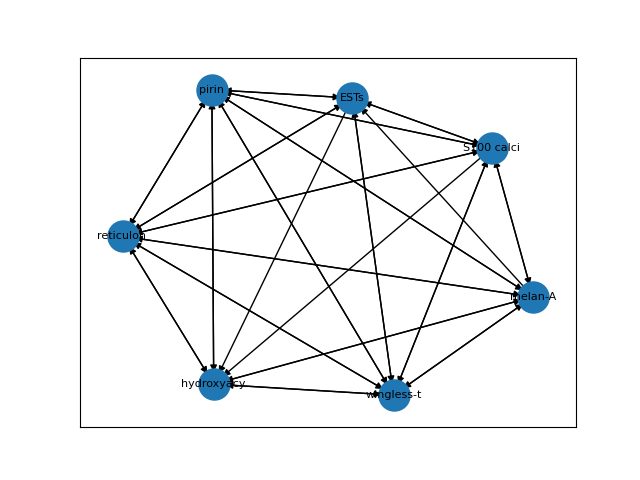}
        \caption{The output is a PBN}
	\end{subfigure}
\end{figure}

In this section we describe the inference method and how it can be implemented. Our approach to inferring a PBN starts with the real gene expression data in the form of a matrix $G$ as input (see Fig. \ref{input-output}(a)), and produces a PBN (see Fig. \ref{input-output}(b)).
The input matrix is of size $m \times n$, where $m$ is the number of genes and $n$ is the number of samples.

The method we apply for inferring PBNs draws upon work done by Shmulevich et al. \cite{smulevich02}.
First, it requires the dataset to be discretised (recall Section \ref{discretisation}).
This process is performed in Algorithm \ref{discAlg}.

\begin{algorithm}[H]
\label{discAlg}
	\SetAlgoLined
	\KwData{Input array G, discretisation axis, calculation method}
	\KwResult{Discretised array G}
	\For{row g in G over axis}{
		t = method(G)\;
		\For{element in g}{
			\eIf{element $<$ t}{
				element = 0\;}
			{
				element = 1\;
				}
		}
	return G \;
	}
\caption{Discretisation algorithm}
\end{algorithm}

Given a target gene, $n_p$ sets of genes with the highest CODs are found.
This is done following Algorithm \ref{predList}.

\begin{algorithm}[H]\label{predList}
	\SetAlgoLined
	\KwData{Input array G, number of input genes per tuple k, number of input tuples n}
	\KwResult{List of predictor-COD-input tuples}
	tuples = []\;
	\For{targetIndex in $1 \dots$ G.geneAxisSize}{
		indexCombinations = generateAllCombinations($1 \dots G.geneAxisSize \setminus targetIndex$, k)\; \tcp{Would return triplets of all possible k-combinations of the gene indexes within the array.}
		buffer.init(n)\;
		\For{inputCombo in indexCombinations}{
			COD, weights = calcCOD(G[targetIndex],G[inputCombo])\;
			\If{COD $>$ min(buffer.COD)}{
				buffer.add(COD, weights, inputCombo)\;
				buffer.removeSmallestCOD() \;
				}
			}
		tuples[targetIndex] $\gets$ buffer\;
		}
	return buffer\;
\caption{Tuple list generator}
\end{algorithm}

A buffer of size $n_p$ is initialised, and each possible combination of input genes have their CODs calculated.
If a combination of inputs has a COD higher than at least one saved in the buffer, the buffer entry with the lowest COD gets replaced by the new combination of inputs.
This results in a buffer full of input combinations with the highest CODs.
One such buffer is initialised per target gene, resulting in $n_p$ input combinations per target gene.

During run-time, a set of input genes is chosen with probability proportional to the COD of the set, and the next state is governed by the state of those input genes in conjunction with the predictive model that was saved.
For all intents and purposes, the list with input gene, perceptron weights and probabilities are enough to construct a PBN, as the input genes convey the connectivity, and the perceptron weights convey the logic for that set of input genes.
The process is summarised in Algorithm \ref{generalAlg}.

\begin{algorithm}[H]\label{generalAlg}
	\SetAlgoLined
	\KwData{Input array G, discretisation axis, threshold calculation method, number of input genes per tuple k, number of input tuples n}
	\KwResult{List of predictor-probability-input tuples}
	$\hat{G}$ = discretise(G, axis, method)\;
	predictorList = genPredictorList($\hat{G}$, k, n)\;
	CODsum = 0 \;
	\For{tuple in predictor list}{
		CODsum $\gets$ CODsum + tuple[COD] \;
		}

	\For{tuple in predictor list}{
		tuple[probability] $\gets \frac{tuple[COD]}{CODsum}$ \;
		}
	return predictor list\;
\caption{General algorithm for generating predictor lists}
\end{algorithm}

\section{Analysis}
\label{analysis}

The analysis of the generated PBNs in our approach are based on steady-state distribution, which is fairly standard, e.g., see \cite{kim03}.
The PBN is run for $T$ steps in order to get it within a steady state. 
Then it is run for the next $N$ steps, recording the state it is at.
To confirm whether or not the PBN is in a steady state after $T$ steps, the Kolmogorov-Smirnov (KS) statistic is calculated for the two halves of $N$.

The entries recorded in $N$ are split in to two halves - one containing states $[0, \frac{N}{2}]$, the other containing $[\frac{N}{2}+1, N]$.
The entries are subsampled with the interval $G$.
The histograms are converted to cumulative distribution functions, and the maximum vertical distance between them is found, which is the KS statistic.

The significance test shows the probability of the two CDFs being drawn from the same distribution.
If the PBN had not reached a steady state after $T$ steps, the halves of $N$ would be drawn from different distributions, which would be indicated by the KS test.
The recorded states are a string of binary values.
Therefore, for ease of analysis, they are used as gray-coded integers, and displayed on a histogram (cf. Fig. \ref{figResults}).
This makes the horizontal distance on the histogram proportional to the Hamming distance between two network states.

\section{Evaluation}\label{eval}

We have implemented the pipeline using Python $3$ and made it 
publicly available on \url{https://github.com/UoS-PLCCN/pbn-inference}.

We have constructed PBNs of size 7 from data produced by Bittner et al. \cite{bittner00} using different thresholds for the quantisation methods.
The thresholds were (a) average of a gene expression; (b) median of a gene expression, and (c) k-means clustering of a gene expression. 
The data was quantised on a per-gene basis, with each gene having 10 triplets of input genes.

\begin{figure}
	\centering
	\caption{
	SSDs of PBNs generated using different quantisation methods.\\ States on the x-axis; SSD probability on the y-axis}\label{figResults}
	\begin{subfigure}[t]{.32\linewidth}
		\includegraphics[width=\linewidth]{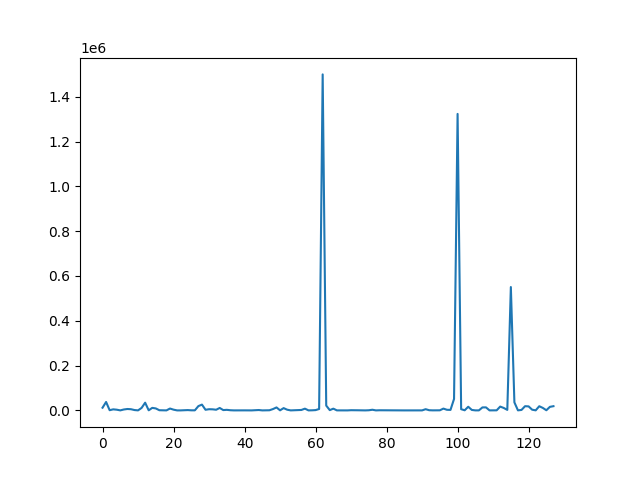}
		\caption{Mean}
	\end{subfigure}
	\begin{subfigure}[t]{.32\linewidth}
		\includegraphics[width=\linewidth]{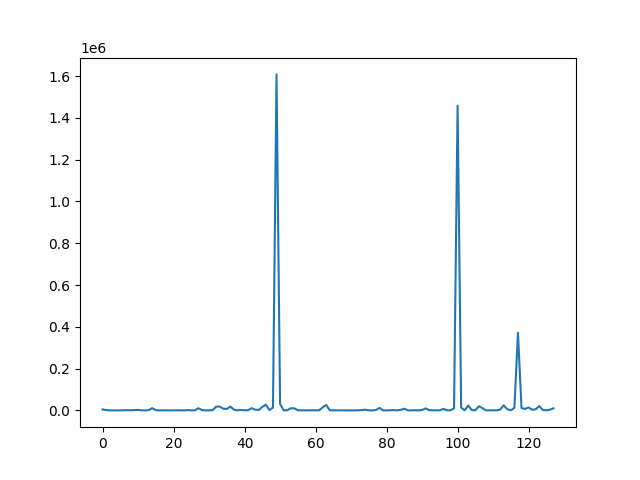}
		\caption{Median}
	\end{subfigure}
	\begin{subfigure}[t]{.32\linewidth}
		\includegraphics[width=\linewidth]{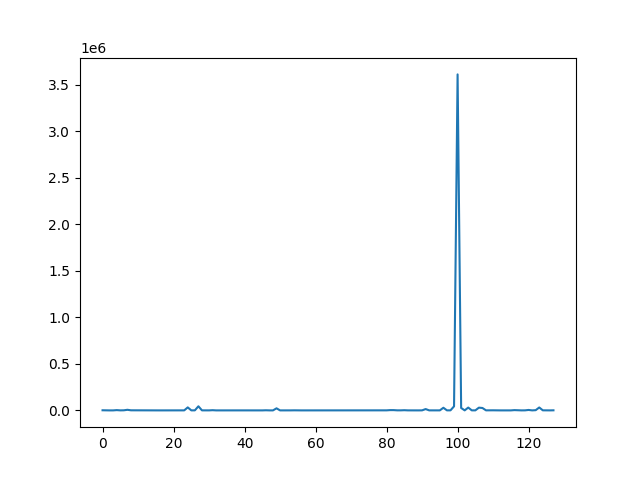}
		\caption{K-Means clustering}
	\end{subfigure}
\end{figure}

For the construction and validation of the histograms representing the steady-state distribution, we have chosen the parameters to be $T = 10^6$, $N=4 \cdot 10^6$, $G=10$ and $R = 100$. 
On a laptop with 32 GB of RAM and an Intel® Core™ i7-7700HQ Processor, each histogram took around 9 hours to produce.
The results are shown in Fig. \ref{figResults}.

It can be seen that the average and the median quantisation methods produce very similar histograms, with three peaks each, and the latter two peaks being in similar positions.
The histogram generated using the PBN constructed from k-means clustering only has one prominent state, which can also be observed in the other two PBNs.
It may be constructive to note that the few very prominent states in the histograms show in Fig. \ref{figResults} agrees with the assumption claimed by Kim et al. \cite{kim03} that gene regulatory networks found in nature only occupy a small fraction of the possible state space.

For the purposes of direct comparison, we have trialled the proposed method in the DREAM (Dialogue on Reverse Engineering Assessment and Methods) challenge\footnote {http://dreamchallenges.org/project/dream-3-in-silico-network-challenge/} which offers a benchmark for network inference (DREAM 3) \cite{marbach-2010-DREAM} and scored 8th (out of 29).

\section{Conclusion}\label{conclusion}

In this work we described the inference a PBN directly from real gene data, collected using microrarray technology, which were taken when the system was at a steady-state. This kind of gene profiling is typically less costly to obtain than time series data, and includes more data points. 
Using the evaluation methods described in the literature, e.g., by Kim et al. \cite{kim03}, we have concluded that the pipeline works well for the examples provided. However, it is subject to fine-tuning the parameters. We have provided the method in a systematic pipeline which can be reproduced and made it publicly available on github \url{https://github.com/UoS-PLCCN/pbn-inference}.

We note that the method scored 8th (out of 29) in the DREAM challenge and has been used to infer large PBNs (N=200). 

It is worth noting is that the proposed method does not require a state transition probability matrix to be produced. It can be extracted from the PBN, however, the time required grows exponentially with the size of the PBN. This means that conventional mathematical methods in the literature that make use of the transition probability matrix may not always be applicable.

One concern is that the transitions 
get fitted to the quantised dataset.
It is widely accepted that the states observed in the dataset are steady states of the cells.
Since the transition rules get fitted to the steady states of the cells, the resulting PBN will be driven towards the steady states observed within the data.
However, while it is certain that the method captures the long-run behaviour (steady-state) of the underlying gene regulatory network, there is little certainty that the PBN will behave with biological accuracy between the observed steady states.
This concern could possibly be addressed by using time-series gene data to augment the method presented here, as this type of data captures the change of gene expression levels with respect to time. This promises to capture the behaviour at and between steady states, without reconstruction of the state evolution of the PBN, and is certainly worth exploring further in future work.


\bibliographystyle{unsrt}
\bibliography{ref}
%
%

\end{document}